\documentclass[a4paper,11pt]{article}
\usepackage{amsmath}
\usepackage{graphicx}
\usepackage{cite}

\begin{document}

\section*{\textbf{A compact tunable polarized X-ray source based on laser-plasma helical undulators}}
~\par

\section*{\textbf{Authors:}}
J. Luo$^{1,2}$, M. Chen$^{1,2\ast}$, M. Zeng$^{1,2}$, J. Vieira$^{3}$, L.L. Yu$^{1,2}$, S.M. Weng$^{1,2}$, L.O. Silva$^{3}$, D.A. Jaroszynski$^{4}$, Z.M. Sheng$^{1,2,4\dag}$, and J. Zhang$^{1,2}$
~\par

\section*{\textbf{Affiliations:}}
$^{1}$Key Laboratory for Laser Plasmas (Ministry of Education) and Department of Physics and Astronomy, Shanghai Jiao Tong University, Shanghai, 200240, China.\\
$^{2}$Collaborative Innovation Center of IFSA (CICIFSA), Shanghai Jiao Tong University, Shanghai 200240, China.\\
$^{3}$GoLP/Instituto de Plasmas e Fusao Nuclear, Instituto Superior Tecnico, Universidade de Lisboa, Lisbon 1049-001, Portugal.\\
$^{4}$SUPA, Department of Physics, University of Strathclyde, Glasgow G4 0NG, UK.\\
$^{\ast}$e-mail: minchen@sjtu.edu.cn\\
${\dag}$e-mail: zhengming.sheng@strath.ac.uk\\
~\par

\section*{\textbf{Abstract:}}
\noindent Laser wakefield accelerators have great potential as the basis for next generation compact radiation sources because their accelerating gradients are three orders of magnitude larger than traditional accelerators. However, X-ray radiation from such devices still lacks of tunability, especially the intensity and polarization distribution. Here we propose a tunable polarized radiation source from a helical plasma undulator based on plasma channel guided wakefield accelerator. When a laser pulse is initially incident with a skew angle relative to the channel axis, the laser and accelerated electrons experience collective spiral motions, which leads to elliptically polarized synchrotron-like radiation with flexible tunability on radiation intensity, spectra and polarization. We demonstrate that a radiation source with millimeter size and peak brilliance of $2\times10^{19}\rm photons/s/mm^{2}/mrad^{2}/0.1\%\ bandwidth$ can be made with moderate laser and electron beam parameters. This brilliance is comparable with the third generation of synchrotron radiation facilities running at similar photon energies, suggesting that laser plasma based radiation sources are promising for advanced applications.
\clearpage

\noindent Tunable x-ray synchrotron radiation sources have wide applications in natural sciences, medicine and industry\cite{Huxley1980N,Chen2014S}. X-rays for such applications are mostly generated from synchrotron radiation facilities, which are usually bulky, costly and not able to satisfy the increasing demand of researchers\cite{Bilderback2005JPBAMOP}. Compact, tunable and flexible light sources are ideal for a broad range of applications and many schemes for such sources are under investigation\cite{Wong2016NPho,Phuoc2005POP}. Among these the laser wakefield accelerator (LWFA) scheme is highly promising due to its tremendous accelerating gradient\cite{Tajima1979PRL,Esarey2009RMP,Hooker2013NPho}. Electrons can be longitudinally accelerated to GeV energy over centimeters\cite{Leemans2014PRL}. Besides longitudinal acceleration, electrons also undergo transverse betatron oscillations in wakefields which lead to the emission of electromagnetic radiation in the x-ray range\cite{Corde2013RMP,Rousse2004PRL,Cipiccia2011NP}. Applications of these radiation sources such as object imaging with high resolution have already been tested\cite{Fourmaux2011OL,Kneip2011APL,Wenz2015NC}. Further applications may become possible if some key properties of these radiation sources, including brightness, spectral features and polarization, can be controlled. Among these the polarization is relatively more difficult to control as it relies on the polarization of electron betatron motion. However, it is usually transversely isotropic. Although the polarization can be controlled through electron injection selection or driver pulse tilting techniques\cite{Schnell2013NC,Popp2010PRL}, the polarization tunability is still not sufficient. Coupling the LWFA with external magnetic undulators would produce tunable radiation. However, both the operation difficulty and the device size increase. External undulators and beam transport systems would enlarge the size of the devices from centimetres to several meters\cite{Schlenvoigt2008NP}.\\
~\par

\noindent Making a more compact radiation source with both small acceleration and radiation lengths is challenging. In order to reach this goal, many kinds of plasma based undulators with small period length have been suggested as possible candidates to replace conventional magnetic undulators\cite{Joshi1987IEEEJQE,Reitsma2008IEEETPS,Andriyash2014NC,Chen2016LSA,Rykovanov2015RPL,Vieira2016arXiv,Hooker2012Patent}. Here, we provide a pathway that may strongly advance compact plasma based undulator technology, by proposing a compact plasma based helical undulator to generate tunable polarized x-ray radiation, thus paving the way to controlled x-ray generation. Through three dimensional particle-in-cell (3D-PIC) simulations we show that the motion of the laser inside the plasma channel depends on the initial incidence offset and angle between the laser and channel axis. We find conditions where the laser spirals about the channel axis, thereby leading to helical betatron trajectories of accelerated electrons. A post radiation calculation code was then used to study the elliptically shaped hollow x-ray emission in far field. Besides the intensity distribution, we find that the polarization and spectra of the radiation are easily controllable by changing the incidence angle of the driver pulse and channel profiles. Considering typical laser and plasma parameters, currently available in many laboratories, we find that our centimeter scale, plasma-based radiation source is expected to deliver ultra-intense x-rays, with peak brilliance as large as $2\times10^{19}\rm photons/s/mm^{2}/mrad^{2}/0.1\%\ bandwidth$. Such high quality flexible and compact radiation source has the potential to impact a wide variety of applications.\\
~\par

\section*{\textbf{Laser and electron beam spiral motions inside the plasma channel}}
\noindent We first study the electron motion inside the plasma channel which resembles that in a helical undulator. Stable laser wakefield accelerators usually use plasma channels to guide the laser pulse over many Rayleigh lengths (i.e. $Z_{R}=k_0w_0^2/2$, where $k_0=2\pi/\lambda_{0}$ is the laser wave number in vacuum, $w_0$ is the focal spot size and $\lambda_0$ is the laser wavelength). A plasma channel with a transverse parabolic density profile of $n(r)=n_0+\Delta nr^2/r_0^2$ can guide the laser for long distances, where $r$ represents the transverse coordinate, $n_0$ is the on-axis electron density, $\Delta n$ is the channel depth and $r_0$ is the channel width\cite{Esarey2009RMP,Chen2016LSA,Rykovanov2015RPL}. For a matched plasma channel (i.e., $\Delta n = \Delta n_c = (\pi r_e r_0^2)^{-1}$ and $r_0=w_0$, where $r_e=e^2/m_{e}c^2$ is the classical electron radius with $e$ and $m_e$ the electron charge and mass at rest, respectively, and $c$ the light speed in vacuum), the laser spot size remains approximately constant during all propagation. If a laser pulse is initially off-axially injected into the plasma channel or at some injection angle, the laser centroid will oscillate transversely in a single plane as it propagates forward\cite{Chen2016LSA}. The oscillation period length, given by $\Lambda_{os} = 2\pi Z_M(\Delta n_c/ \Delta n)^{1/2}$, where $Z_M=\pi r_0^2/\lambda_0$, can be controlled by tuning the channel depth\cite{Chen2016LSA}. In addition, when the laser wave number vector is non-collinear with the channel axis, the laser centroid will perform a spiral trajectory around the channel axis as shown in Fig. 1(a). The centroid trajectory of the laser pulse can be simply approximated from the eikonal equations according to the ray tracing method\cite{Born1975Book} (details can be found in Supplementary Information). In cylindrical coordinates ($r,\ \phi,\ z$), the laser centroid trajectory is then given by
\begin{eqnarray}
\frac{dr}{dz}&=&\pm \frac{1}{cos\theta_z} [1-\frac{n(r)}{n_c}-(1-\frac{n_{e0}}{n_c})\cdot(sin^2\theta_z \frac{b^2}{r^2}+cos^{2}\theta_z)] ^{1/2} \\
\frac{d\phi}{dz}&=&\frac{bsin\theta_z}{r^2cos\theta_z}
\end{eqnarray}
with $\theta_z = arccos(\sqrt{1-cos^{2}\theta_x-cos^{2}\theta_y})$, and where $\theta_{x,\ y,\ z}$ are the angles between the laser propagation direction and the space coordinate axes x, y, and z. Moreover, $n_c$ is the critical density of plasma, $n_{e0}=n_{e}(r=\sqrt{x_0^2+y_0^2})$ is the plasma density at the laser incidence point ($x_0$, $y_0$). The parameter $b=(y_0-x_{0}cos\theta_y/cos\theta_x)/(\sqrt{(cos\theta_y/cos\theta_x)^2+1})$ is called striking distance, i.e. the distance between the projection of the incident laser pulse and the plasma channel centre.\\
~\par

\noindent When an ultrashort ultraintense laser pulse propagates in such a plasma channel, the background electrons are expelled outward due to the pondermotive force of the driver pulse, generating bubble like wakefields at the back of the laser pulse\cite{Pukhov2002APB,Lu2007PRSTAB}. If the laser pulse oscillates transversely, the bubble will follow the laser, forcing electrons inside the bubble to follow the motion of the laser centroid\cite{Reitsma2008IEEETPS}. The oscillation period and amplitude can be kept constant until near the laser pump depletion\cite{Chen2016LSA}. Thus the plasma may act as a helical undulator and the radiation of the electrons shows typical helical undulator radiation properties, which are dramatically different from those of normal betatron radiation.\\
~\par

\noindent We used the massively parallel, fully-relativistic PIC code OSIRIS\cite{Fonseca2002LNCS} to investigate the self-consistent three-dimensional dynamics of the laser centroid motion and electron acceleration inside the plasma channel when the laser wave vector is not aligned with the channel axis in the strongly nonlinear blowout regime. Detailed simulation parameters can be found in the Methods. The channel axis is set along the z axis ($x=y=0$). In a typical simulation shown in Fig. 1, the initial parameters of laser incidence are $x_0=1\ \mu m$, $y_0=0$, $\theta_x=89^{\circ}$ and $\theta_y=91^{\circ}$. Typical bubble cross sections at different laser propagation distances are plotted in Fig. 1(b) to 1(e). Figure 2(a) shows a typical snapshot of the three-dimensional structure of the laser driven blowout region (yellow), with the externally injected electrons at the back of the bubble (light green), and the laser electric fields (blue-red-orange-dark green). Projections of the laser propagation direction on x-y and x-z planes (black dashed arrows) in Fig. 2(a) and a movie of the laser and wake evolution (refer to the Supplementary Information) indicate the spiral motion. The black solid lines in Fig. 2(b) show typical trajectories of the laser centroid obtained from theoretical approximations by solving Eqs. (1) and (2). The red solid lines represent the trajectories from 3D-PIC simulations. The dashed lines on the side planes are the corresponding trajectory projections. The analytical model correctly predicts the wavelength and amplitude of the laser centroid oscillations, both in excellent agreement with PIC simulations. These results show that the eikonal equations can be used to describe the laser propagation for the typical propagation distances in our investigation. The oscillation wavelength from the PIC simulation is about $700\ \mu m$ closed to the theoretical approximation $\Lambda_{os}=719.5\ \mu m$. In the simulations, because of the laser energy depletion and frequency red shifting as the laser pulse propagates further, the oscillation wavelength and amplitude become smaller than theoretical predictions.\\
~\par

\noindent To reduce simulation time and concentrate on the study of the properties of the radiation emitted by the trapped electrons, we consider an external injected scenario, employing a cylindrical electron beam to study the radiation dependance on the intial laser incidence. An illustration of a simulation including ionization injection is given in the Supplementary Information, in order to illustrate the feasibility of our scheme in an all optical setup by only using LWFA. As mentioned before, the trapped and accelerated electron beam experiences spiral motion simutaneously with the acceleration structure following the laser pulse inside the plasma channel. We randomly traced 200 particles from the accelerated electron beam. The centroid position of these electrons ($x_c,\ y_c,\ z_c$) are shown in Fig. 2(b) by the blue solid line. The Root-Mean-Square radius of these electrons ($r_{rms}=\sqrt{<(x_i-x_c)^2+(y_i-y_c)^2>}$) is less than $0.63\ \mu m$ during the whole simulation process. As one can see from Fig. 2(b), the period of the spiral motion of the electrons closely follows the period of the laser centroid oscillations. In addition, the amplitude of the oscillations is slightly higher than the laser centroid oscillations, due to its inertia. Electrons are accelerated to nearly 140 MeV in only $1600\ \mu m$. The rms deviation of the average energy increases to about 8 MeV, which results from the nonumiformity of the longitudinal acceleration fields in the helically moving bubble. The continous acceleration of electrons with a stable accelerating gradient about $80\ \rm{GeV}$$/m$ before $x=1600\ \mu m$ illustrates that electrons propagate near the back of the bubble. After $x>1600\ \mu m$, the accelerating gradient reduces rapidly as electrons enter the dephasing region. So the following calculation of electrons trajectories and radiation stops at $x=1600\ \mu m$, which corresponds to using a short capillary in practice.\\
~\par

\section*{\textbf{X-ray radiation from the helical plasma undulator}}
\noindent The spiral motion of the trapped electrons inside the wakefield resembles the electron motion in a traditional helical undulator, which is common in a storage ring of a synchrotron facility for polarized x-ray emission\cite{Wang1994NIMIPRA}. We studied the radiation properties of these electrons with the numerical radiation post-processing code VDSR\cite{Chen2013PRSTAB} (see Methods). This code calculates the radiation distribution according to electrons¡¯ trajectories obtained from the 3D-PIC OSIRIS simulations. At the beginning of the electron acceleration the normal beam betatron period of the electrons inside the accelerating bubble ($\lambda_{\beta}\approx\sqrt{2\gamma}\lambda_p$) is far smaller than the laser oscillation period ($\Lambda_{os}$), the radiation properties are more relied on the betatron motion and the emitted photons are low energies which are not the ones we are interested in. In the radiation calculations we cut out these low energy radiation part. Figure 3(a) shows the radiation intensity distribution captured by a virtual radiation detector located in the far-field and centred along the plasma channel axis (see schematic view in Fig. 1(a)). The initial injection parameters of the laser are the same as those in Fig. 1. The far field radiation pattern illustrated in Fig. 3(a) shows a hollow elliptical shape, which is consistent with the electron trajectories whose projection on x-y plane is shown in Fig. 2(b). Key properties of the radiation intensity distribution are described by a fundamental dimensionless radiation parameter, called the strength parameter, given by $K=2\pi\gamma r_{os}/\Lambda_{os}$, where $\gamma$ is the Lorentz factor of the electrons, $r_{os}$ is their transverse oscillation amplitude and $\Lambda_{os}$ is the spiral motion period. In this case the electrons are rapidly accelerated to high energies ($\gamma\approx274$). Moreover, the minimum oscillation amplitude is $(r_{os})_{min}\approx1.41\ \mu m$, the maximum oscillation amplitude is $(r_{os})_{max}\approx6.84\ \mu m$, and $\Lambda_{os}\approx700\ \mu m$. These parameters leads to a minimum strength parameter $K_{min}\approx3.5$ and a maximum strength parameter $K_{max}\approx16.8$. Both $K_{min}$ and $K_{max}$ are larger than unity, which means the radiation is in the wiggler regime. As a result, the radiation spectrum is broadband. The x-rays propagate along the emission angles $\theta\sim K/\gamma\approx0.73^{\circ}$ and $3.51^{\circ}$. These angles are larger than the angular width around the emission angle $\Delta\sim 1/\gamma\approx0.21^{\circ}$, therefore leading to the hollow intensity pattern shown in Fig. 3(a).\\
~\par

\noindent To investigate the spectral properties, we have selected two points with the highest radiation intensities and marked them with black circles. As one can see, the maximum radiation intensity comes from the major axis vertices of the radiation ellipse with radiation angle about $\theta\approx3.5^{\circ}$, which corresponds to the positions in the electron trajectory projection plane where the electrons have maximum deviation ($r_{os}\approx(r_{os})_{max}$) and experience the maximum transverse acceleration. The photon energy spectra of the marked points are shown in Figs. 3(c) and 3(d). The peak position of the spectrum radiated along this direction is about 400 eV which is about the average value of the on-axis radiation where the peak radiation energy is $E_p=\frac{3}{2}K\gamma^2hc/\Lambda_{os}=3\pi\gamma^3r_{os}hc/\Lambda_{os}^2$ with $h$ representing the Planck¡¯s constant.\cite{Corde2013RMP} For average electron energy ($\bar{\gamma}=137$) it is $E_p=418.9 eV$.\\
~\par

\noindent In order to discuss the polarization properties of the radiated photons, we define the polarization direction degree of a point as $P=I_x/I_t$, where $I_x$ is the radiation intensity with photon polarization in the x direction and $I_t$ is the total radiation intensity. When $P=1$ the radiation is linearly polarized in the x direction, while when $P=0$ the radiation is linearly polarized in the y direction. The polarization direction degree distribution is shown in Fig. 3(b). The two points marked in Fig. 3(a) were also lined out. During the spiral motion, the directions of the electrons' transverse velocities keep changing, and so does the radiation polarization direction. Because of the positive correlation between radiation polarization and electrons¡¯ velocity direction, the polarization direction degree appears to change periodically along the elliptical orbit. When electrons move to the upper right marked position, its velocity gradient component in the x direction is lower than that in the y direction, so $P<0.5$ and the radiation intensity in the x direction (corresponding to red line in Fig. 3(c)) is weaker than that in the y direction (blue line in Fig. 3(c)). The opposite situation at the lower left marked position is shown in Fig. 3(d).\\
~\par

\noindent In addition to the polarization, the radiation intensity is also another important parameter for useful photon sources. The radiation intensity is proportional to the total number of electrons in the beam for incoherent radiation emission. In our typical 3D-PIC simulations, due to the limitation of computational resources, small laser focal spot sizes and injected electron beam radii have been used. The externally injected electron beam charge used in the simulation is thus 4.5 pC, which is close to the ionization induced injection charge measured in the simulation reported in the Supplementary Information. With this charge, the totally number of radiated photons is about $2\times10^7$. A much larger photon number would be radiated if the total number of trapped electrons could be higher. In general, higher charges can be accelerated when using higher energy drivers propagating in lower density plasma in comparison to those considered in our work. Higher accelerated beam charge could also be obtained in ionization injection scenarios by increasing the concentration of the injection gas component. Considering the electron rotation transverse area $15.1\ \mu m^2$ and the bunch duration 6.67 fs, in our case the final maximum radiation brightness is about $2\times10^{19}\rm photons/s/mm^{2}/mrad^{2}/0.1\%\ bandwidth$, which is comparable with the highest levels of current third generation of synchrotron radiation facilities at similar x-ray energies.\\
~\par

\noindent In our scheme the intensity and polarization distributions of the radiation can be easily controlled by varying the initial laser incidence parameters, i.e. the initial off-axis position ($x_0,\ y_0$) and the tilted injection angles ($\theta_x,\ \theta_y$). In Fig. 4(a) the laser centroid trajectories with different initial parameters are shown. The black line corresponds to the one shown in Fig. 2. The radiation shows hollow elliptical shape. An additional simulation was also performed in order to obtain a hollow and circular intensity profile by using laser injection parameters of $x_0=2.92\ \mu m$, $y_0=0$, and $\theta_x=90^{\circ}$, $\theta_y=91.5^{\circ}$. The laser performs circular rotation around the channel axis initially. Near laser pump depletion, however, radius of rotation decreases as shown by the red line in Fig. 4(a). Because the emission angle is larger than the angular width of the radiation, the radiation intensity distribution also exhibits a hollow circle pattern, illustrated in Fig. 4(b). The corresponding radiation polarization distribution is shown in Fig. 4(c). As one can see the polarization is strongly angle dependent and it is along the azimuthal direction. This kind of distribution is very convenient for many light-polarization-dependent applications.\\
~\par

\noindent Quasi linearly polarized x-rays can also be generated by selecting appropriate initial laser incidence. If one fixes the initial incidence position ($x_0=1\ \mu m$, $y_0=0$), and incidence angle $\theta_y=91^{\circ}$, but only decreases $\theta_x$  from $89^{\circ}$ to $88^{\circ}$, the ellipticity decrease of the laser centroid trajectory is evident from the blue line in Fig. 4(a). Correspondingly the electron radiation intensity distribution shows a linear structure (shown in Fig. 4(e)) and a bigger radiation polarization degree $P$ is obtained (see Fig. 4(f)), which means that most of the radiated photons are linearly polarized. Furthermore, in this simulation the oscillation amplitude of the electrons has been increased as one can see in Fig. 4(a). The strength parameter ($K$) is then increased as well. We find that with these parameters the final peak photon energy moves to 1.5 keV, which shows the tunability of radiation energy. Another way to tune the radiation energy is by varying the channel depth since the strength parameter ($K$) depends on the laser oscillation period ($\Lambda_{os}$) in the parabolic plasma channel. The effects of the channel depth on the radiation parameters has already been studied in detail in Refs. 22 and 23.\\
~\par

\noindent In general, fine tunability of the distribution of the radiation intensity and polarization can be obtained by controlling the combinations of the initial laser incidence parameters. Here we only show the tunability of the elliptical azimuth angle of the radiation pattern by controlling the combined parameter of $cos(\theta_x)/cos(\theta_y)$ with fixed incidence position ($x_0,\ y_0$). The results are shown in Fig. 4(d). The radiation tilt angle can be tuned from 20 to 70 degrees freely. Because of the coincidence of the electron motion and the laser centroid motion, in practice, one can design the radiation properties by solving the eikonal equations of the laser motion first.\\
~\par

\section*{\textbf{Conclusion}}
\noindent In conclusion, we have studied highly tunable soft x-ray radiation from a helical plasma undulator based on LWFA in a plasma channel. When the initial laser pulse propagation axis is in an arbitrary skew angle relative to the channel axis, theoretical and simulation results illustrate that both the laser pulse and electron beams carry out spiral motion. The plasma acts as a helical undulator modulating electron motion, such that elliptically polarized synchrotron-like radiation can be produced. The distributions of the radiation intensity, polarization and spectrum can all be tuned by controlling the initial laser incidence parameters and the profile of the plasma channel. In general x-ray radiations with hollow elliptical shape and elliptical polarization with brightness comparable to or even higher than the third generation of synchrotron radiation sources can be obtained from these laser plasma based helical undulators. Our studies reveal that such all optical-plasma compact radiation sources can be made highly flexible and are therefore suitable for advanced applications. It deserves to point out that although currently there are still many factors such as average photon flux, spectrum width, stability, making wakefield based radiation sources uncompetitive with traditional SRs, the uniqueness of such sources ( for example brightness, compactness and synchronization with high power laser) make them beneficial complements of SRs. In further, the recent progresses on wakefield acceleration (i.e. electron injection\cite{Zeng2015PRL}, wake diagnosis\cite{Matlis2006NP} and high repetition lasers\cite{Mourou2013NPho}) have shown the potential to make more stable, high quality and high flux radiation sources.\\
~\par

\section*{\textbf{Methods}}
\noindent 3D-PIC simulations were performed with OSIRIS code. In the simulations, the profile of the normalized laser vector potential is given by $a=eA/m_ec^2=a_0\times exp(-t^2/L_0^2-r^2/w_0^2)$ with $a_0=2.0$ (corresponding to the peak intensity of the polarized laser pulse of $I=8.6\times10^{18}\ W/cm^2$), $L_0=4.0\ T_0$, $w_0=6.75\ \lambda_0$ and the laser wavelength $\lambda_0=0.8\ \mu m$, the laser period $T_0=2\pi/\omega_0\approx2.67\ \rm fs$. The plasma density has a parabolic profile with $n_0=0.001\ n_c$, $\Delta n=\Delta n_c$, $r_0=w_0$, where $n_c\approx1.7\times10^{21}\ cm^{-3}$ is the critical plasma density for the driver pulse. The injected external electron beam has transverse and longitudinal dimensions with size of $L_r=0.5\ \mu m$, $L_z=2.0\ \mu m$, and a charge of $Q\approx4.5\ \rm pC$. The energy of the beam is $E_e=7.7\ \rm MeV$ or 5.1 MeV in different simulation cases. The size of the simulation box is $40\times40\times45\ \mu m^3$ corresponding to cell size of $200\times200\times1035$. We use 4 super-particles per cell and 8 super-particles per cell for the background electrons and external injected electrons, respectively.\\
~\par

\noindent Post process code VDSR was used to perform the radiation calculation. The spatial and velocity vectors of 200 super-particles randomly selected from the accelerated electron beam in 3D-PIC simulations were input into the VDSR code. The time internal of trajectory recording is $dt=0.04\ T_0$. The code integrates each particle's radiation along its trajectory. Since the electron beam size is much longer compared with the radiation wavelength interested here, the radiation is essentially incoherent. Because of the computational source limitation, the number of the traced particle is too small for a coherent addition of the electron radiation, we then sum them incoherently to get the final far field spectra on a virtual screen with spatial resolutions. The detected circular region covers a maximum polar angle of $6^{\circ}$, which is divided into 24 parts in the radial direction and 72 parts in the azimuthal direction.\\
~\par

\section*{\textbf{Author contributions}}
\indent M.C. conceived the idea; J.L., M.C. and M.Z. performed the PIC simulations and radiation calculations; J.L., M.C., Z.M.S., J.V. analyzed data; J.L., M.C., Z.M.S. and J.Z. wrote the main manuscript. J.V., L.L.Y., S.M.W., L.O.S. and D.A.J. improved the manuscript. All authors commented on the manuscript and agreed on the contents.\\
~\par

\section*{\textbf{Competing financial interests}}
\indent The authors declare no competing financial interests.\\
~\par

\section*{\textbf{Acknowledgements}}
\indent This work is supported in part by the National Basic Research Program of China (2013CBA01504), the National Science Foundation of China (11421064, 11374209, 11374210), a MOST international collaboration project (2014DFG02330), and an EU Horizon 2020 Programme project EuPRAXIA (Grant No. 653782). M.C. appreciates supports from National 1000 Young Talent Program. Z.M.S. acknowledges the support of a Leverhulme Trust Research Grant. J.V. and L.O.S. acknowledge Laserlab \uppercase\expandafter{\romannumeral3} (Contract No. 284464) and Laserlab \uppercase\expandafter{\romannumeral4} (Contract No. 654148). D.A.J. acknowledges support of the U.K. EPSRC (Grant No. EP/J018171/1), the EC's LASERLAB-EUROPE (Grant No. 654148), EuCARD-2 (Grant No. 312453). The authors would like to acknowledge the OSIRIS Consortium for the use of OSIRIS and the visXD framework. Simulations were performed on the $\rm \Pi$ supercomputer at Shanghai Jiao Tong University and Tianhe \uppercase\expandafter{\romannumeral2} supercomputer at Guangzhou.\\

\clearpage
\section*{\textbf{Figures}}

\begin{center}
\includegraphics[width=1\textwidth]{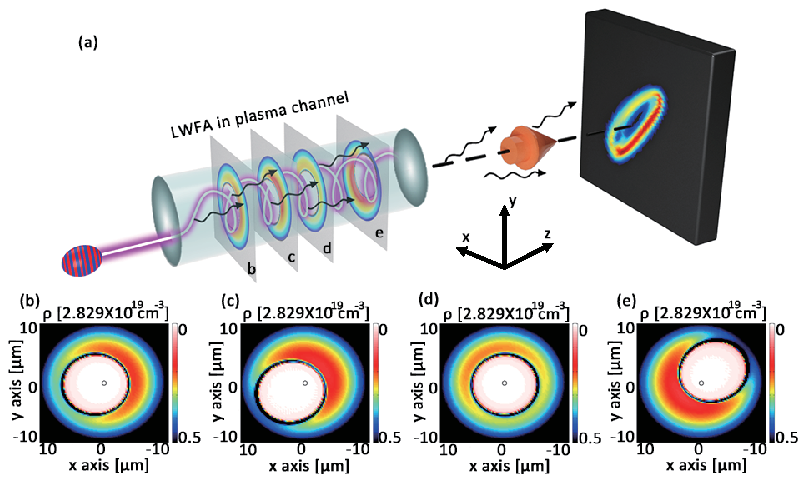}
\end{center}
\small \noindent Fig. 1 \textbf{Sketch of the radiation source from a helical plasma undulator based on LWFA.} (\textbf{a}) Spiral motion of laser pulse, wakefield in the plasma channel and the far field x-ray radiation distribution. (\textbf{b, c, d, e}) are the electron density distributions of the transverse slices of the accelerating bubble structure with largest radius at t = 133 fs (\textbf{b}), t = 600 fs (\textbf{c}), t = 1200 fs (\textbf{d}), t = 1800 fs (\textbf{e}).\\
~\par
~\par
~\par

\begin{center}
\includegraphics[width=1\textwidth]{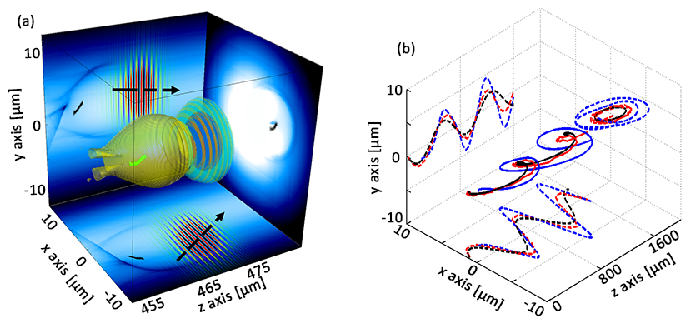}
\end{center}
\small \noindent Fig. 2 \textbf{Spiral motions of the laser pulse and electron beam in the plasma channel.} (\textbf{a}) A snapshot of plasma density (blue background), bubble structure (yellow surface), injected electron beam (light green points), and laser electric fields (blue-red-orange-green iso-surfaces and the projections). The arrows show the laser propagation direction in projection planes. (\textbf{b}) Theoretical (black solid line) and simulation (red solid line) results of the laser pulse centroid trajectory. The blue solid line shows the centroid trajectory of the traced particles. Dashed lines are the projections of the trajectories.\\
~\par
~\par
~\par

\begin{center}
\includegraphics[width=1\textwidth]{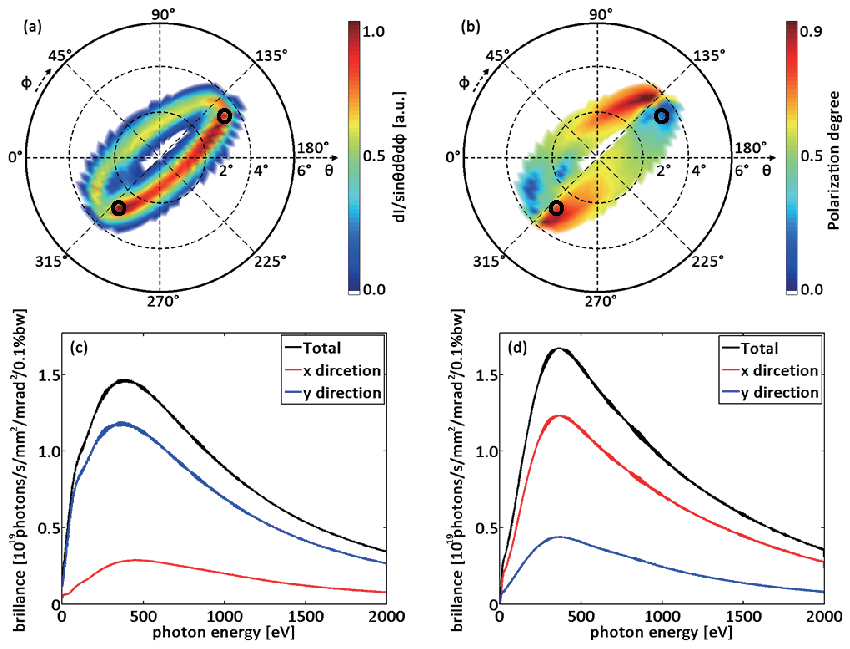}
\end{center}
\small \noindent Fig. 3 \textbf{Intensity and polarization distributions of x-ray radiation.} (\textbf{a}) Far field radiation intensity distribution, (\textbf{b}) polarization distribution, (\textbf{c}) radiation spectrum of the marked upper right point and lower left point (\textbf{d}).\\
~\par
~\par
~\par

\begin{center}
\includegraphics[width=1\textwidth]{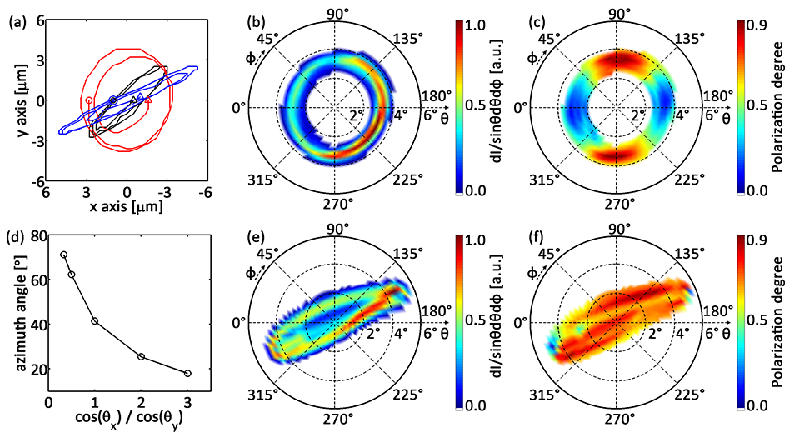}
\end{center}
\small \noindent Fig. 4 \textbf{Radiation distributions of intensity and polarization for different laser incidence parameters.} (\textbf{a}) Laser centroid trajectories on x-y plane with initial laser incidence parameters of $x_0=1\ \mu m$, $\theta_x=89^{\circ}$, $\theta_y=91^{\circ}$ (black), $x_0=2.92\ \mu m$, $\theta_x=90^{\circ}$, $\theta_y=91.5^{\circ}$ (red), and $x_0=1\ \mu m$, $\theta_x=88^{\circ}$, $\theta_y=91^{\circ}$. Input parameter $y_0$ is fixed to be $0\ \mu m$. The circles and triangles mark the start and end positions of these trajectories, respectively. (\textbf{b, e}) are radiation intensity distributions and (\textbf{c, f}) are radiation polarization distributions. (\textbf{b, c}) correspond to the red line in (\textbf{a}), (\textbf{e, f}) correspond to blue line in (\textbf{a}). (\textbf{d}) The correlation between the azimuth angle of the elliptical radiation region and the ratio of the cosine function of incidence parameters $\theta_x$ and $\theta_y$.
\end{document}